\documentclass[aps,prl,twocolumn,superscriptaddress,showpacs,showkeys,amsmath,amssymb,floatfix]{revtex4}
\usepackage{booktabs,graphicx,mathrsfs,verbatim}
\usepackage[colorlinks=true,citecolor=blue,filecolor=blue,linkcolor=blue,urlcolor=blue,pdftex]{hyperref}
\usepackage[usenames,dvipsnames]{color}
\usepackage{ulem}
\usepackage{color}

\newcommand{\1}[1]{\, \mathrm{#1}} 
\newcommand{\n}[1]{\mathrm{#1}} 

\newcommand{\percent}{\%}

\newcommand{\arxiv}[1]{\href{http://arxiv.org/abs/#1}{\texttt{arXiv:#1}}}

\newcommand{\assergi}{\affiliation{INFN, Laboratori Nazionali del Gran Sasso, Assergi, 67100, Italy}}
\newcommand{\bologna}{\affiliation{University of Bologna and INFN-Bologna, Bologna, Italy}}
\newcommand{\columbia}{\affiliation{Physics Department, Columbia University, New York, NY 10027, USA}}
\newcommand{\coimbra}{\affiliation{Department of Physics, University of Coimbra, R. Larga, 3004-516, Coimbra, Portugal}}
\newcommand{\heidelberg}{\affiliation{Max-Planck-Institut f\"ur Kernphysik, Saupfercheckweg 1, 69117 Heidelberg, Germany}}
\newcommand{\houston}{\affiliation{Department of Physics, Rice University, Houston, TX 77005 - 1892, USA}}

\newcommand{\losangeles}{\affiliation{Physics \& Astronomy Department, University of California, Los Angeles, USA}}
\newcommand{\mainz}{\affiliation{Institut f\"ur Physik, Johannes Gutenberg Universit\"at Mainz, 55099 Mainz, Germany}}
\newcommand{\munster}{\affiliation{Institut f\"ur Kernphysik, Wilhelms-Universit\"at M\"unster, 48149 M\"unster, Germany}}
\newcommand{\shanghai}{\affiliation{Department of Physics, Shanghai Jiao Tong University, Shanghai, 200240, China}}
\newcommand{\subatech}{\affiliation{SUBATECH, Ecole des Mines de Nantes, CNRS/In2p3, Universit\'e de Nantes, 44307 Nantes, France}}
\newcommand{\weizmann}{\affiliation{Department of Particle Physics and Astrophysics, Weizmann Institute of Science, 76100 Rehovot, Israel}}
\newcommand{\zurich}{\affiliation{Physics Institute, University of Z\"{u}rich, Winterthurerstr. 190, CH-8057, Switzerland}}

\begin{document}

\title{Dark Matter Results from 100 Live Days of XENON100 Data}

\author{E.~Aprile}\columbia
\author{K.~Arisaka}\losangeles
\author{F.~Arneodo}\assergi
\author{A.~Askin}\zurich
\author{L.~Baudis}\zurich
\author{A.~Behrens}\zurich
\author{K.~Bokeloh}\munster
\author{E.~Brown}\munster
\author{T.~Bruch}\zurich
\author{G.~Bruno}\assergi
\author{J.~M.~R.~Cardoso}\coimbra
\author{W.-T.~Chen}\subatech
\author{B.~Choi}\columbia
\author{D.~Cline}\losangeles
\author{E.~Duchovni}\weizmann
\author{S.~Fattori}\mainz
\author{A.~D.~Ferella}\zurich
\author{F.~Gao}\shanghai
\author{K.-L.~Giboni}\columbia
\author{E.~Gross}\weizmann
\author{A.~Kish}\zurich
\author{C.~W.~Lam}\losangeles
\author{J.~Lamblin}\subatech
\author{R.~F.~Lang}\email{rafael.lang@astro.columbia.edu}\columbia
\author{C.~Levy}\munster
\author{K.~E.~Lim}\columbia
\author{Q.~Lin}\shanghai
\author{S.~Lindemann}\heidelberg
\author{M.~Lindner}\heidelberg
\author{J.~A.~M.~Lopes}\coimbra
\author{K.~Lung}\losangeles
\author{T.~Marrod\'an~Undagoitia}\zurich
\author{Y.~Mei}\houston\mainz
\author{A.~J.~Melgarejo~Fernandez}\columbia
\author{K.~Ni}\shanghai
\author{U.~Oberlack}\houston\mainz
\author{S.~E.~A.~Orrigo}\coimbra
\author{E.~Pantic}\losangeles
\author{R.~Persiani}\bologna
\author{G.~Plante}\columbia
\author{A.~C.~C.~Ribeiro}\coimbra
\author{R.~Santorelli}\zurich
\author{J.~M.~F.~dos Santos}\coimbra
\author{G.~Sartorelli}\bologna
\author{M.~Schumann}\email{marc.schumann@physik.uzh.ch}\zurich
\author{M.~Selvi}\bologna
\author{P.~Shagin}\houston
\author{H.~Simgen}\heidelberg
\author{A.~Teymourian}\losangeles
\author{D.~Thers}\subatech
\author{O.~Vitells}\weizmann
\author{H.~Wang}\losangeles
\author{M.~Weber}\heidelberg
\author{C.~Weinheimer}\munster

\collaboration{The XENON100 Collaboration}\noaffiliation


\begin{abstract}
We present results from the direct search for dark matter with the XENON100 detector, installed underground at the Laboratori Nazionali del Gran Sasso of INFN, Italy. XENON100 is a two-phase time projection chamber with a 62~kg liquid xenon target. Interaction vertex reconstruction in three dimensions with millimeter precision allows the selection of only the innermost 48~kg as ultra-low background fiducial target. In 100.9 live days of data, acquired between January and June 2010, no evidence for dark matter is found. Three candidate events were observed in the signal region with an expected background of $(1.8\pm0.6)$~events. This leads to the most stringent limit on dark matter interactions today, excluding spin-independent elastic WIMP-nucleon scattering cross-sections above $7.0\times10^{-45}\1{cm^2}$ for a WIMP mass of $50\1{GeV/c^2}$ at 90\% confidence level.
\end{abstract}

\pacs{
 95.35.+d, 
 14.80.Ly, 
 29.40.-n, 
}

\keywords{Dark Matter, Direct Detection, Xenon}

\maketitle


Weakly Interacting Massive Particles (WIMPs) are a well-motivated class of particles~\cite{Steigman:1984ac;Jungman:1995df} to constitute a major fraction of the dark matter in the Universe~\cite{Jarosik:2010iu;Nakamura:2010zzi}. These particles can be searched for in underground-based detectors~\cite{Goodman:1984dc} through their coherent scattering off target nuclei. A quasi-exponentially falling energy spectrum of nuclear recoils is expected, extending up to a few tens of keV at most for WIMPs scattering off a xenon target~\cite{Lewin:1995rx}. A wide range of WIMP-nucleon cross sections predicted by theoretical models remains untested by current-generation searches, with the most stringent limits on the elastic spin-independent WIMP-nucleon cross-section coming from CDMS-II~\cite{Ahmed:2009zw}, EDELWEISS-II~\cite{Armengaud:2011cy} and XENON100~\cite{Aprile:2010um}.


XENON100 is the current phase of the XENON dark matter program, which aims to improve the sensitivity to dark matter interactions in liquid xenon (LXe) with two-phase (liquid/gas) time-projection chambers (TPCs) of large mass and low background. A key feature of XENON100 is its ability to localize events with millimeter resolution in all spatial dimensions, enabling the selection of a fiducial volume in which the radioactive background is minimized. The simultaneous detection of charge and light signals provides discrimination between the expected WIMP-induced nuclear recoil (NR) signal and interactions from the electromagnetic background in the form of electronic recoils (ERs). First results from XENON100~\cite{Aprile:2010um} have shown a sensitivity competitive with that from the full 612~kg$\times$days exposure of CDMS-II \cite{Ahmed:2009zw} after only 11.2~live days of data taking. Here, the result of a WIMP search using 100.9~live days of XENON100 data is reported.


XENON100 is filled with 161~kg of ultra-pure LXe. Of these, 99~kg are used as active scintillator veto, surrounding the optically separated 62~kg target in $4 \pi$. Thanks to careful material selection~\cite{Aprile:2011ru} and detector design, XENON100 achieves an experimentally verified ER background in the relevant low-energy region of \mbox{$<5\times10^{-3}$~events$\times$(keV$_{\n{ee}}\times$kg$\times$day)$^{-1}$} (keV$_{\n{ee}}=$ keV electron-equivalent~\cite{Aprile:2008rc}) before signal discrimination~\cite{Aprile:2011vb}. An interaction in the cylindrical LXe target of $\sim$30~cm height and $\sim$30~cm diameter generates prompt scintillation light~(S1) and ionization electrons, the latter being detected through the process of proportional scintillation~(S2) in the gaseous xenon above the liquid. Both S1 and S2 signals are registered by photomultiplier tubes (PMTs), at the bottom of the LXe target for optimal light collection, and placed above in the gas phase. The interaction vertex is reconstructed in 3~dimensions, with the $(x,y)$-position determined from the hit pattern of the localized S2 signal on the top PMT array, and the $z$-coordinate deduced from the drift time between the S1 and S2 signals. This allows the fiducialization of the target volume to exploit the excellent self-shielding capabilities of LXe. Due to their different ionization densities, ERs ($\gamma$, $\beta$ background) and NRs (WIMP signal or neutron background) have a different S2/S1 ratio, which is used as discrimination parameter.


The 242 PMTs used in XENON100 are $1''$-square Hamamatsu R8520-AL PMTs with a quantum efficiency of $\sim$30\% at the Xe light wavelength of 178~nm, and low intrinsic radioactivity~\cite{Aprile:2011ru}. The measured average energy threshold of the LXe veto is $\sim100\1{keV_{ee}}$.


The TPC is installed inside a vacuum insulated stainless steel cryostat which is surrounded by a passive shield made of high purity copper, polyethylene, lead and water in order to suppress external backgrounds. A constant flow of high-purity nitrogen boil-off gas keeps the $^{222}$Rn level inside the shield $<1\1{Bq/m^3}$. A 200~W pulse tube refrigerator, installed outside the shield structure, keeps the detector at its operating temperature of $-91^\circ$C, with excellent stability over time (fluctuations $<$0.05\%). To bring calibration sources ($^{60}$Co, $^{137}$Cs, $^{241}$AmBe) close to the target, a copper tube penetrates the shield and winds around the cryostat. XENON100 is installed underground at the Italian Laboratori Nazionali del Gran Sasso (LNGS) below an average $3600\1{m}$ water equivalent rock overburden, which reduces the muon flux by a factor~$\sim10^6$.


At low energies, the event trigger is provided by the S2 signal. The summed signal of 84~central PMTs is shaped and fed into a low-threshold discriminator. The trigger efficiency has been measured to be $>99\%$ at 300~photoelectrons (PE) in S2.


Three algorithms are used to reconstruct the $(x,y)$~coordinates of the events. They yield consistent results out to a radius of 14.2~cm, with the active TPC radius being 15.3~cm. The $(x,y)$~resolution was measured with a collimated source and is $<$3~mm~($1\sigma$). The algorithm based on a Neural Network gives the most homogeneous response and thus is used for event positioning, while the information from the other algorithms is used for consistency checks. The drift time measurement gives a $z$-position resolution of $0.3\1{mm}$ ($1\sigma$) and allows to distinguish of two interaction vertices if separated by more than 3~mm in~$z$. The positions are corrected for non-uniformities of the drift field, as inferred from a finite-element simulation and validated by data.


XENON100 uses continuous xenon purification through a hot getter. The mean electron lifetime $\tau_e$ is indicative of the amount of charge lost to impurities~\cite{Aprile:2009dv}. It increased from $230\1{\mu s}$ to $380\1{\mu s}$ for the data reported here, as measured weekly with $^{137}$Cs calibrations. A linear fit to the $\tau_e$ time evolution yields the $z$-correction for the S2~signals with negligible systematic uncertainty~($<2.5\%$). $(x,y)$ variations of the S2 signal are corrected using a map obtained with the $662\1{keV_{ee}}$ line from $^{137}$Cs.


The spatial dependence of the S1~signal due to the non-uniform light collection is corrected for using a map obtained with the $40\1{keV_{ee}}$ line from neutrons scattering inelastically on $^{129}$Xe. It agrees within 3\% with maps inferred from data using the $662\1{keV_{ee}}$ line and the $164\1{keV_{ee}}$ line, from neutron-activated $^{131m}$Xe. The light yield $L_y(122\1{keV_{ee}})=(2.20\pm0.09)\1{PE/keV_{ee}}$ at the applied drift field of $530\1{V/cm}$ in the LXe is determined by a fit to the light yields measured with all available calibration lines~\cite{Aprile:2010um}.


\begin{figure}[htbp]
\centering\includegraphics[width=1\columnwidth]{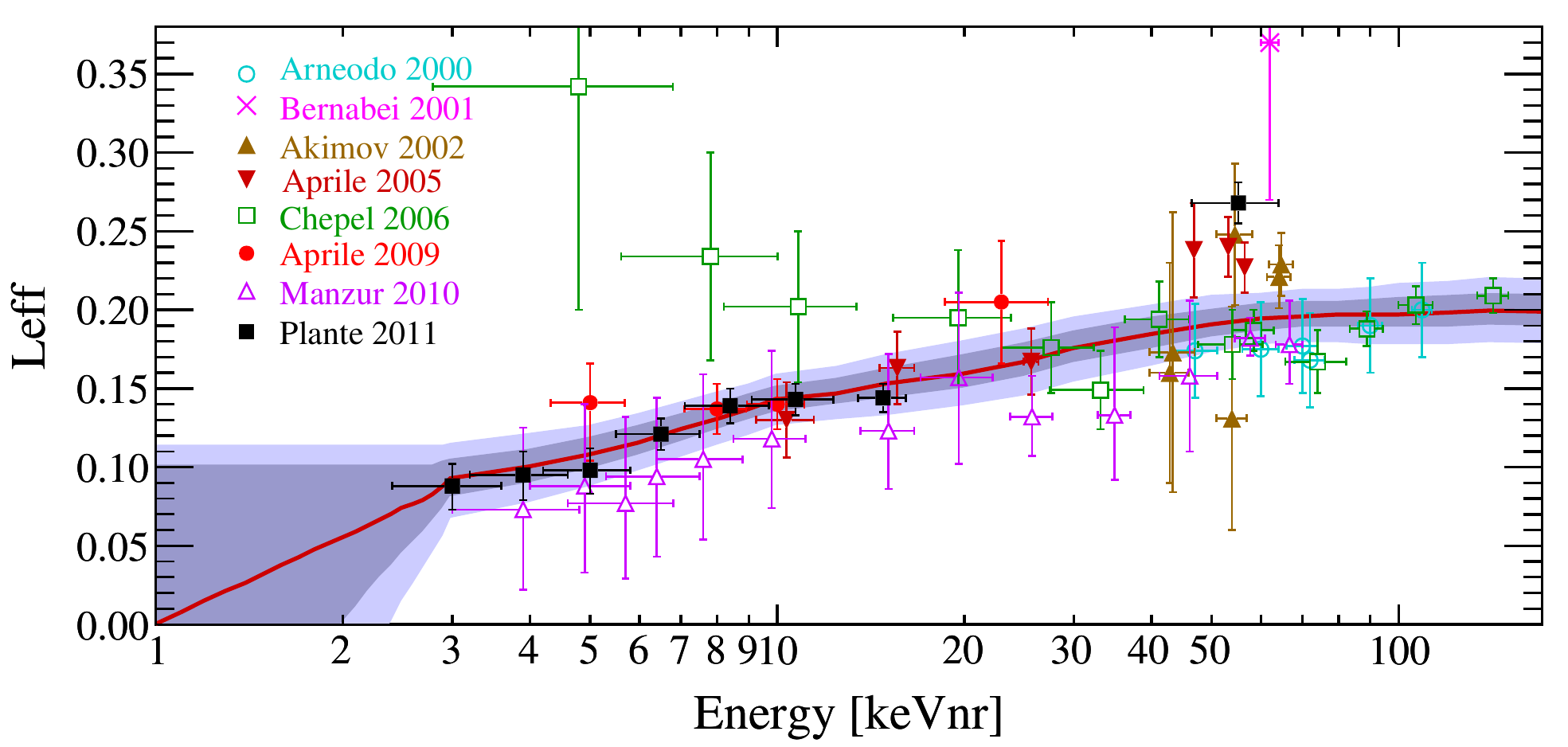}
\caption{All direct measurements of $\mathcal{L}_{\text{eff}}$ \cite{Plante:2011,Arneodo:2000vc}
described by a Gaussian distribution to obtain the mean (solid line) and the uncertainty band
($1\sigma$ and $2\sigma$). Below $3\1{keV_{nr}}$
the trend is logarithmically extrapolated to $\mathcal{L}_{\text{eff}}=0$ at $1\1{keV_{nr}}$.}\label{fig:leff}
\end{figure}

The NR energy $E_{\n{nr}}$ is inferred from the S1 signal using $E_{\n{nr}} \! = \! (S1/L_y) (1/\mathcal{L}_{\text{eff}})(S_{\n{ee}}/S_{\n{nr}})$. The scintillation efficiency $\mathcal{L}_{\text{eff}}$ of NRs relative to that of $122\1{keV_{ee}}$ $\gamma$-rays at zero field is taken from the parametrization shown in Fig.~\ref{fig:leff}, which is strongly supported by the most recent measurements by the Columbia group~\cite{Plante:2011} but includes all direct measurements of this quantity~\cite{Arneodo:2000vc}.
$\mathcal{L}_{\text{eff}}$ is logarithmically extrapolated below the lowest measured energy of $3\1{keV_{nr}}$, motivated by the trend in the data points as well as studies which simultaneously take into account light and charge signal~\cite{Bezrukov:2010qa}. The electric field scintillation quenching factors for ERs $S_{\n{ee}}=0.58$ and NRs $S_{\n{nr}}=0.95$ are taken from~\cite{Aprile:2006kx}.

The dark matter data presented here was acquired between January~13 and June~8, 2010. About 2\%~of the exposure was rejected due to variations in detector operation parameters. In addition, 18~live days of data taken in April were rejected due to an increased electronic noise level. With an average data acquisition live time of $\sim90\%$, and regular calibration runs during the data-taking period, this leads to a dataset of 100.9~live days. This data had been blinded below the 90\% ER quantile in $\log_{10}$(S2/S1)-space for S1$<$160~PE.

It was decided a-priori to derive the dark matter result based on a Profile Likelihood analysis as introduced in~\cite{Aprile:2011hx} but taking into account all relevant backgrounds for this dataset. This analysis does not use a cut based on the S2/S1 discrimination. Both, the signal and the background hypothesis would be tested regardless of the observed data. In parallel, an analysis based on the optimum interval method~\cite{Yellin:2002xd} was also performed. The restricted S2/S1 space used in the latter analysis, together with the energy interval, defines a benchmark WIMP search region that allows to directly compare the observed signal with the expected background.

From a comparison of the measured background rate with Monte Carlo simulations of the XENON100 electromagnetic background~\cite{Aprile:2011vb}, a $^{\n{nat}}$Kr concentration of $(700\pm100)$~ppt is inferred for the data reported here, higher than
reported earlier~\cite{Aprile:2010um}. The additional Kr was introduced by an air leak during maintenance work on the gas recirculation pump, prior to the start of the data-taking period. This results in an expected ER background of $<22\times10^{-3}$~evts$\times(\mathrm{keV_{ee}\!\times \!kg\!\times\! day)}^{-1}$ before S2/S1 discrimination, more than an order of magnitude lower than the one of other dark matter search experiments.
After the science run presented here, the Kr concentration in the Xe has been reduced by cryogenic distillation to the level reported in~\cite{Aprile:2010um}, as confirmed with a $\beta$-$\gamma$-coincidence method~\cite{Alimonti:2000xc}. XENON100 is taking new data with this reduced background and improved performance.

\begin{figure}[htbp]
\centering
\includegraphics[width=1\columnwidth]{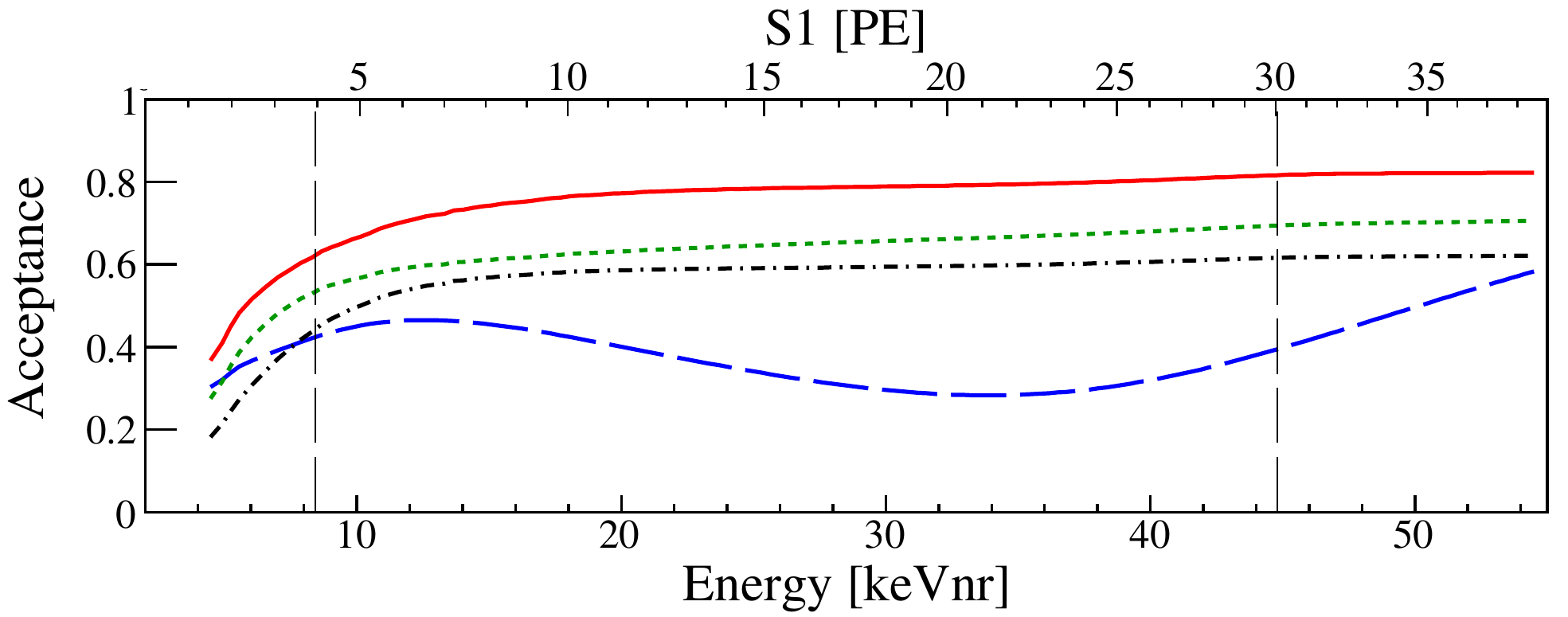}
\caption{Acceptance of all data quality cuts used for the analysis for $m_\chi \ge 50$~GeV/$c^2$ (solid red), $m_\chi = 10$~GeV/$c^2$ (dotted green), $m_\chi = 7$~GeV/$c^2$ (dash-dotted black). The optimum interval analysis additionally uses a S2/S1 ER discrimination cut. Its NR acceptance is also shown (dashed blue).}\label{fig:acceptance}
\end{figure}

The imposed requirements to the quality and topology of events are designed to retain the highest possible acceptance of the expected WIMP-induced single-scatter NRs. The majority of cuts were designed and fixed before unblinding the signal region, based on expected signal characteristics, on NR data from a calibration of 2.9~live days with a $^{241}$AmBe source, and on low-energy ERs from Compton-scattered gammas from a $^{60}$Co source, which were recorded over a total of 5.8~live days spread in time. To reject interactions with a very high energy deposition, the identified S1 and S2 signals are required to amount for more than half of the total digitized signal. To satisfy the primary requirement of the WIMP signature to be a single scatter interaction, one S2~signal above 300~PE is required, corresponding to about 15~ionization electrons. All other S2~signals have to be small enough to be consistent with PMT afterpulsing or delayed ionization signals from the same single-scatter interaction. The corresponding S1~signal must be above 4 PE and must satisfy a two-fold PMT coincidence in a $\pm$20~ns window, without having a coincident signal in the LXe veto. Any other S1-like signal must be consistent with electronic noise or unrelated to the S2, based on its S2/S1 ratio. In addition, both S1 and S2 PMT hit patterns as well as the width of the S2 pulse are required to be consistent with a single interaction vertex at the reconstructed position. The cumulative cut acceptance, used by both analyses, is shown in Fig.~\ref{fig:acceptance} and has an error of~$\sim3\%$. It is estimated based on Monte Carlo simulations, $^{241}$AmBe and $^{60}$Co calibration data, as well as ERs recorded outside the WIMP search region during the dark matter search. It includes a WIMP mass $m_\chi$ dependent S2 acceptance which is derived from the expected recoil spectrum and the measured S2 vs.~S1 distribution.

The energy window for the WIMP search is chosen between $4-30\1{PE}$, corresponding to $8.4-44.6\1{keV_{nr}}$ based on the $\mathcal{L}_{\text{eff}}$ parametrization shown in Fig.~\ref{fig:leff}. The lower bound was assumed to give a sufficient discrimination between genuine S1~signals and electronic noise, whereas the upper bound was chosen to include most of the expected WIMP signal.
To discriminate ERs from NRs, the parameter $\log_{10}(\n{S2_b/S1})$ is used, with $\n{S2_b}$ being the sum of the bottom PMT signals. This choice is motivated by the more uniform signal distribution on these, giving a smaller uncertainty from the $(x,y)$ S2-correction.
To maximize the sensitivity given the homogeneously distributed $^{85}$Kr background, the fiducial volume was optimized on ER background data and set to 48~kg,
and the ER rejection level was set to 99.75\%. The acceptance to NRs below this line is calculated based on single-scatter NRs from $^{241}$AmBe data and is also shown in Fig.~\ref{fig:acceptance}.
The Profile Likelihood approach tests the full S2/S1 space and does not employ this cut.


The expected background in the WIMP search region is the sum of Gaussian leakage from the ER background, of non-Gaussian leakage, and of NRs from neutron interactions. The latter, estimated by Monte Carlo simulations, takes into account neutron spectra and total production rates from $(\alpha,n)$ and spontaneous fission reactions in the detector and shield materials, and is based on the measured radioactivity concentrations~\cite{Aprile:2011ru}. The impact of muon-induced neutrons is obtained from simulations and contributes 70\% to the total. Taking into account the measured trigger efficiency and the energy threshold in the active LXe veto, the overall prediction is $(0.31^{+0.22}_{-0.11})$~single scatter NRs in the 100.9~days data sample before a S2/S1-cut, in the energy region of interest and 48~kg fiducial mass, of which $(0.11^{+0.08}_{-0.04})$ are expected in the benchmark WIMP search region.

\begin{figure}[htbp]
\centering
\includegraphics[width=1\columnwidth]{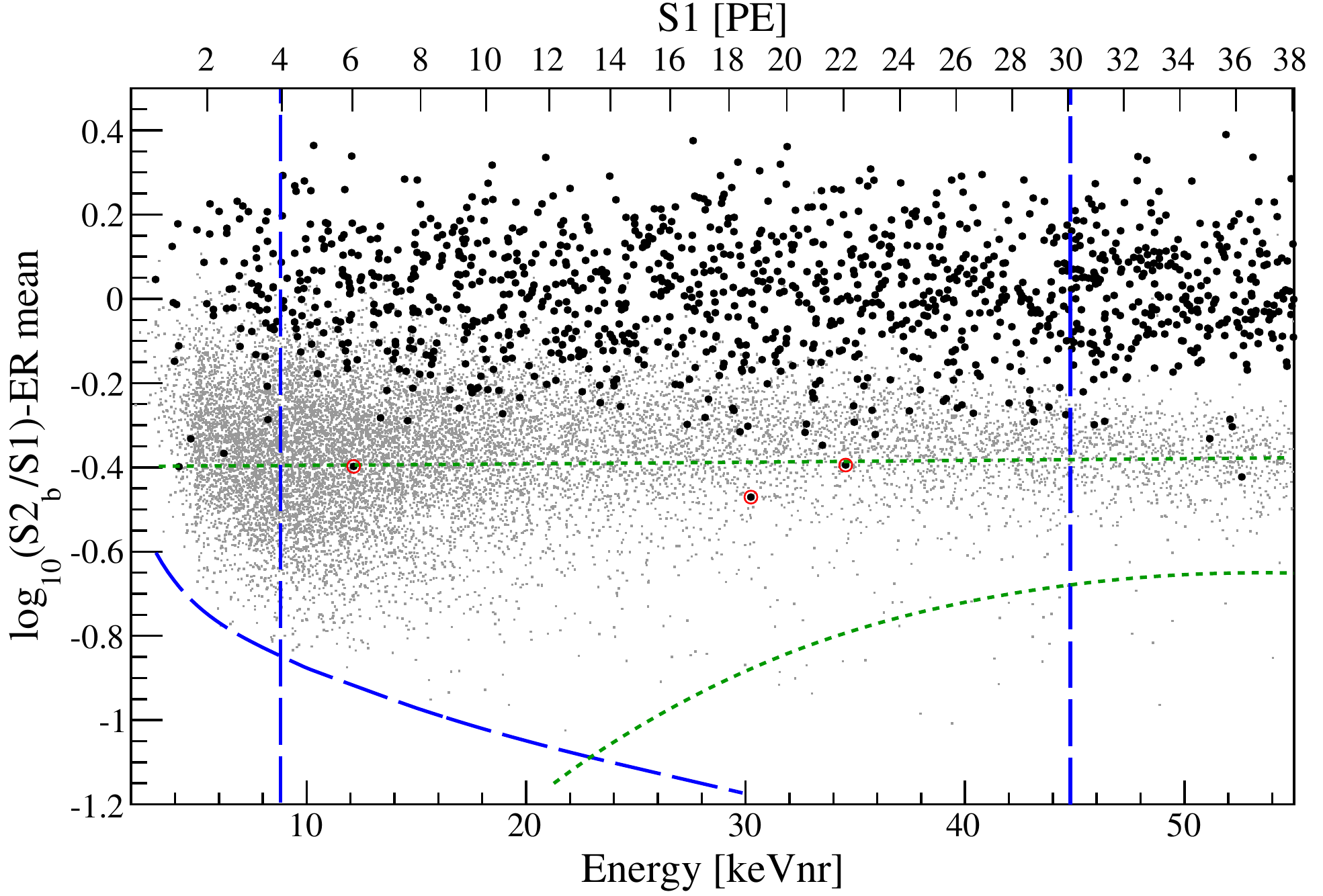}
\caption{Observed event distribution using the discrimination parameter $\log_{10}(\n{S2_b/S1})$, flattened by subtracting the ER band mean, as a function of NR equivalent energy ($\n{keV_{nr}}$). All quality cuts, including those defined after unblinding, are used. Gray points indicate the NR distribution as measured with an $^{241}$AmBe neutron source. The WIMP search region is defined by the energy window $8.4-44.6\1{keV_{nr}}$ ($4-30\1{PE}$) and the lower bound of the software threshold $\n{S2}>300\1{PE}$ (blue dashed). The optimum interval analysis additionally uses the 99.75\% rejection line from above and the $3\sigma$ contour of the NR distribution from below (green dotted). Three events fall into this WIMP search region (red circles), with $(1.8 \pm 0.6)$ events expected from background.}\label{fig:scatter}
\end{figure}


The normalized ER band, obtained by subtracting its mean as inferred from $^{60}$Co calibration data, is well described by a Gaussian distribution in $\log_{10}(\n{S2_b/S1})$ space. Gaussian leakage, dominated by the $^{85}$Kr background, is predicted from the number of background events outside the blinded WIMP search region, taking into account the blinding cut efficiency and the ER rejection level. It is $(1.14 \pm 0.48)$~events in the benchmark WIMP search region, where the error is dominated by the statistical uncertainty in the definition of the discrimination line. Non-Gaussian (anomalous) leakage can be due to double-scatter gamma events with one interaction in a charge insensitive region, e.g.~below the cathode, and one in the active target volume. Such events have a lower effective S2/S1 ratio, since only one interaction contributes to the S2, but both to the S1. Their contribution has been estimated using $^{60}$Co calibration data, taking into account the different exposure compared to background data, and accounting for the fact that the background of this data set is dominated by $^{85}$Kr which $\beta$-decays and does not contribute to such event topologies. The spatial distribution of leakage events for background and calibration data is similar within 10\%. This is verified by Monte Carlo simulations and by data, selecting potential leakage candidates by their S1~PMT hit pattern. Anomalous leakage is estimated to give $(0.56^{+0.21}_{-0.27})$~events, where the uncertainty takes into account the difference in the background and calibration distributions, and that the leakage might be overestimated because of the uncertainty in the $^{85}$Kr concentration. In summary, the total background prediction in the WIMP search region for 99.75\% ER rejection, 100.9~days of exposure and 48~kg fiducial mass is $(1.8\pm0.6)$~events. This expectation was verified by unblinding the high energy sideband from $30-130\1{PE}$ before unblinding the WIMP search region. The Profile Likelihood analysis employs the same data and background assumptions to obtain the prediction for Gaussian, non-Gaussian and neutron background for every point in the $\log_{10}(\n{S2_b/S1})$ parameter space.


After unblinding the pre-defined WIMP search region, a population of events was observed that passed the S1 coincidence requirement only because of correlated electronic noise that is picked up from an external 100~kHz source, as verified by inspection of the digitized PMT signals. These events are mostly found below the S1 analysis threshold, with 3~events from this population leaking into the WIMP search region close to the 4~PE lower bound. This population can be identified and rejected with a cut on the S1~PMT coincidence level, that takes into account correlated pick-up noise, and by cutting on the width of the S1~candidate. These post-unblinding cuts have a combined acceptance of 99.75\% for NRs while removing the entire population of noise events.


\begin{figure}[h!]
\centering
\includegraphics[width=1\columnwidth]{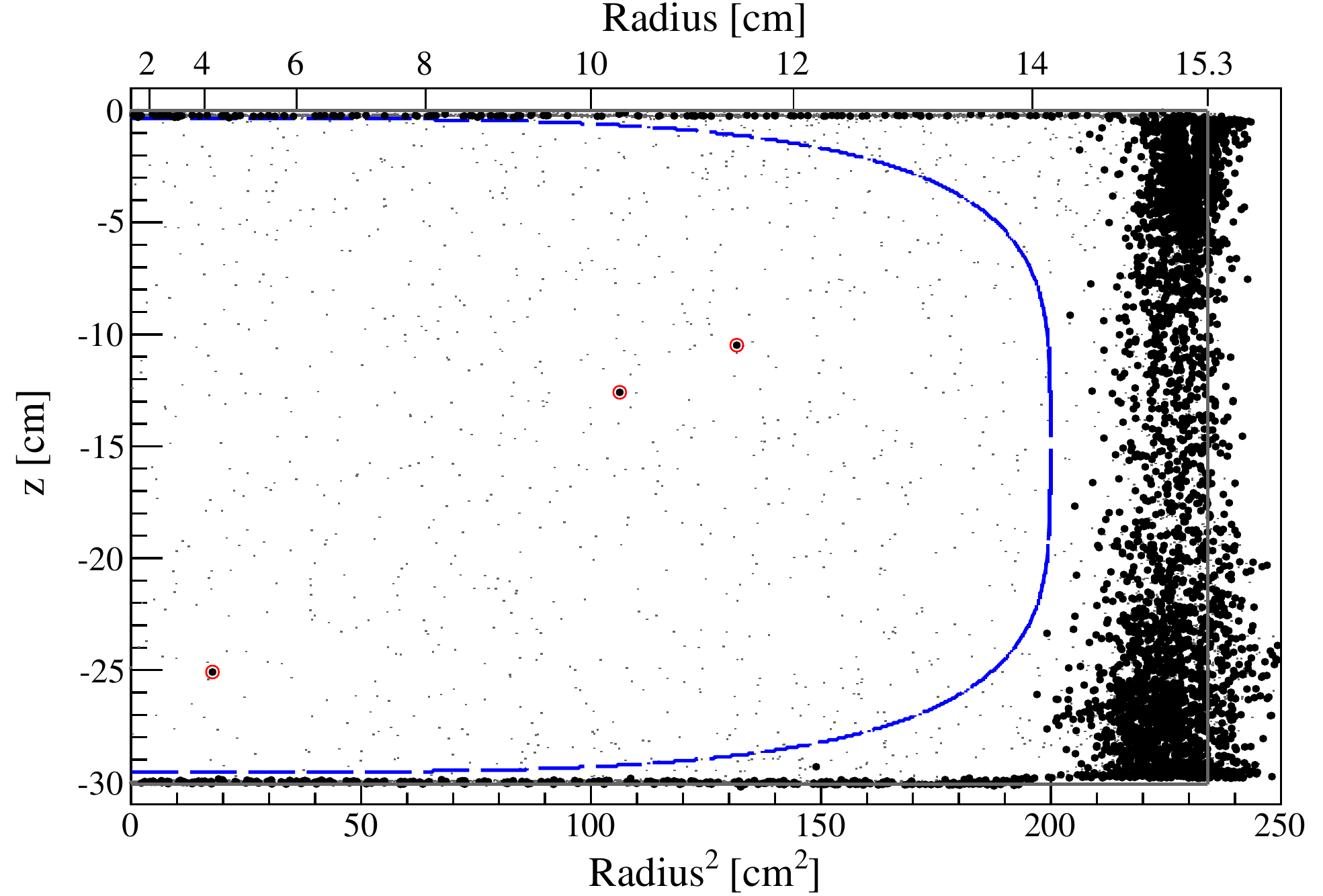}
\caption{Distribution of all events (gray dots) and events below the 99.75\% rejection line (black dots) in the TPC observed in the $8.4-44.6\1{keV_{nr}}$ energy range during 100.9~live days. All cuts are used here, including the ones introduced post-unblinding to remove a population due to electronic noise. The 48~kg fiducial volume (dashed, blue) and the TPC dimensions (gray) are also indicated.}\label{fig:fiducial}
\end{figure}

With these additional cuts, 3~events pass all quality criteria for single-scatter NRs and fall in the WIMP search region, see Fig.~\ref{fig:scatter}. This observation remains unchanged for moderate variations in the definition of any of the data quality cuts. These events were observed on January~23, February~12, and June~3, at $30.2~\1{keV_{nr}}$, $34.6~\1{keV_{nr}}$, and $12.1~\1{keV_{nr}}$, respectively.
The event distribution in the TPC is shown in Fig.~\ref{fig:fiducial}. Given the background expectation of $(1.8\pm0.6)$ events, the observation of 3~events does not constitute evidence for dark matter, as the chance probability of the corresponding Poisson process to result in 3~or more events is 28\%.


\begin{figure}[tbph]
\centering
\includegraphics[width=1\columnwidth]{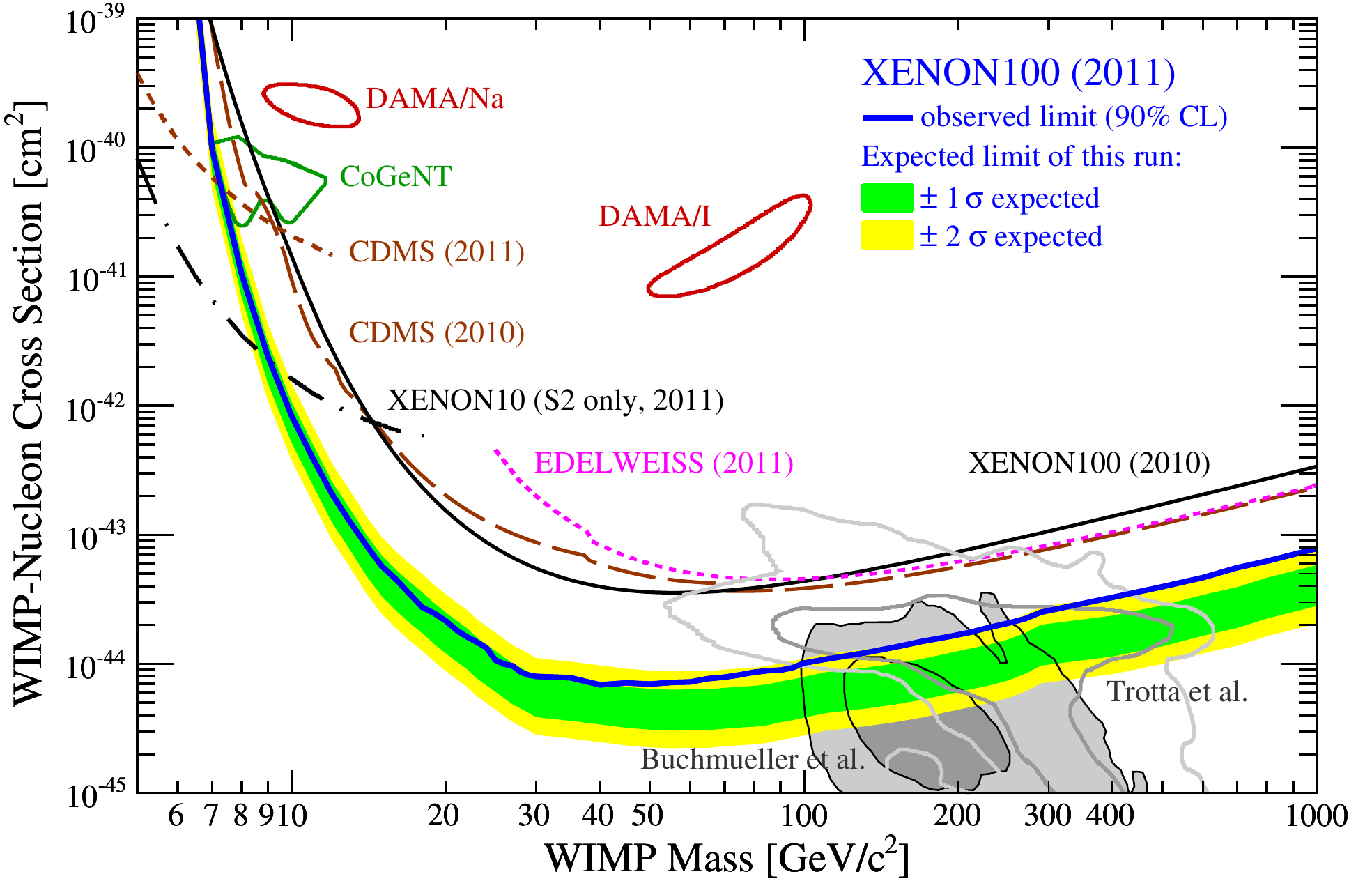}
\caption{Spin-independent elastic WIMP-nucleon cross-section $\sigma$ as function of WIMP mass $m_{\chi}$. The new XENON100 limit at 90\%~CL, as derived with the Profile Likelihood method taking into account all relevant systematic uncertainties, is shown as the thick (blue) line together with the expected 
sensitivity of this run (yellow/green band). The limits from XENON100 (2010)~\cite{Aprile:2010um}, EDELWEISS (2011)~\cite{Armengaud:2011cy}, CDMS (2009)~\cite{Ahmed:2009zw} (recalculated with $v_{\n{esc}}=544$~km/s, $v_0 = 220\1{km/s}$), CDMS (2011)~\cite{Ahmed:2010wy} and XENON10 (2011)~\cite{Angle:2011th} are also shown. Expectations from CMSSM are indicated at 68\% and 95\%~CL (shaded gray~\cite{Buchmueller:2011aa}, gray contour~\cite{Trotta:2008bp}), as well as the $90\percent$~CL areas favored by CoGeNT~\cite{Aalseth:2010vx} and DAMA (no channeling)~\cite{Savage:2008er}.}\label{fig:limit}
\end{figure}

The statistical analysis using the Profile Likelihood method~\cite{Aprile:2011hx} does not yield a significant signal excess either, the $p$-value of the background-only hypothesis is 31\%. A limit on the spin-independent WIMP-nucleon elastic scattering cross-section $\sigma$ is calculated where
WIMPs are assumed to be distributed in an isothermal halo with $v_0=220$~km/s, Galactic escape velocity $v_{\n{esc}}=(544^{+64}_{-46})$~km/s, and a density of $\rho_\chi=0.3\1{GeV/cm^3}$. The S1~energy resolution, governed by Poisson fluctuations of the PE generation in the PMTs, is taken into account. Uncertainties in the energy scale as indicated in Fig.~\ref{fig:leff}, in the background expectation and in $v_{\n{esc}}$ are profiled out and incorporated into the limit. The resulting 90\% confidence level (CL) limit is shown in Fig.~\ref{fig:limit} and has a minimum $\sigma=7.0\times10^{-45}\1{cm^2}$ at a WIMP mass of $m_\chi=50\1{GeV/c^2}$. The impact of $\mathcal{L}_{\text{eff}}$ data below $3\1{keV_{nr}}$ is negligible at $m_\chi=10\1{GeV/c^2}$. The sensitivity is the expected limit in absence of a signal above background and is also shown in Fig.~\ref{fig:limit}.
Due to the presence of two events around $30\1{keV_{nr}}$, the limit at higher $m_\chi$ is weaker than expected. Within the systematic differences of the methods, this limit is consistent with the one from the optimum interval analysis, which calculates the limit based only on events in the WIMP search region. Its acceptance-corrected exposure, weighted with the spectrum of a $m_\chi=100\1{GeV/c^2}$ WIMP, is $1471\1{kg\times days}$. This result excludes a large fraction of previously unexplored WIMP parameter space, and cuts into the region where supersymmetric WIMP dark matter is accessible by the LHC~\cite{Buchmueller:2011aa}. Moreover, the new result challenges the interpretation of the DAMA~\cite{Savage:2008er} and CoGeNT~\cite{Aalseth:2010vx} results as being due to light mass WIMPs.


We gratefully acknowledge support from NSF, DOE, SNF, Volkswagen Foundation, FCT, R\'egion des Pays de la Loire, STCSM, DFG, and the Weizmann Institute of Science. We are grateful to LNGS for hosting and supporting XENON.

\vspace{-.8cm}



\end{document}